\documentclass[a4paper,10pt,twoside]{cpc-hepnp}

\usepackage{multicol}
\usepackage{graphicx}
\usepackage{booktabs}
\usepackage{amssymb,bm,mathrsfs,bbm,amscd}
\usepackage[tbtags]{amsmath}
\usepackage{lastpage}
\usepackage{CJK}

\usepackage{epstopdf}
\usepackage{epsfig}

\begin{document}
\begin{CJK*}{GB}{gbsn}

\fancyhead[c]{\small Chinese Physics C~~~Vol. 42, No. 6 (2018)
064102} \fancyfoot[C]{\small 064102-\thepage}

\footnotetext[0]{Received 22 March 2018}

\title{Phenomenological study on the decay widths of $\Upsilon(nS)\rightarrow \bar{d}^\ast(2380)+X$\thanks{Supported by National Natural Sciences Foundations
of China under the grant Nos. 11475186, 11475192, 11521505, and
11565007, the Sino-German CRC 110 "Symmetries and the Emergence of
Structure in QCD" project by NSFC under the grant No.11621131001,
the Key Research Program of Frontier Sciences, CAS, Grant No.
Y7292610K1, and the IHEP Innovation Fund under the grant No.
Y4545190Y2. }}

\author{%
      Chao-Yi L\"{u}$^{1,2;1)}$\email{lvcy@ihep.ac.cn}%
\quad P. Wang$^{2,3;2)}$\email{pwang4@ihep.ac.cn}%
\quad Y. B. Dong$^{2,3}$\\
      P. N. Shen$^{2,3}$
\quad Z. Y. Zhang$^{2,3}$
\quad D. M. Li$^{1}$
}
\maketitle

\address{%
$^1$ Department of Physics, Zhengzhou University, Zhengzhou, Henan 450001, China\\
$^2$ Institute of High Energy Physics, Chinese Academy of Sciences, Beijing 100049, China, China\\
$^3$ Theoretical Physics Center for Science Facilities, Chinese Academy of Sciences, Beijing 100049, China\\
}

\begin{abstract}
The decay widths of $\Upsilon(nS)$ $\rightarrow$
$\bar{d}^{\ast}(2380)+X$ with $n=1,2,3$ are studied in a
phenomenological way. With the help of crossing
symmetry, the decay widths are obtained by investigating the imaginary
part of the forward scattering amplitudes
between $d^\ast$ and $\Upsilon(nS)$. The wave functions of
$d^\ast$ and deuteron obtained in previous studies are used
for calculating the amplitude. The interaction between $d^{\ast}$
($d$) and $\Upsilon$ is governed by the quark-meson
interaction, where the coupling constant is determined by fitting the
observed widths of $\Upsilon(nS)$ $\rightarrow$ $\bar{d}+X$. The
numerical results show that the decay widths of
$\Upsilon(nS)$ $\rightarrow$ $\bar{d}^{\ast}+X$ are about
$2 - 10$ times smaller than that of
$\bar{d}+X$. The calculated momentum of $\bar{d^*}$ is in the range $0.3-0.8$ GeV.
Therefore, it is very likely that one can find
$\bar{d}^\ast(2380)$ in these semi-inclusive decay processes.
\end{abstract}



\footnotetext[0]{\hspace*{-3mm}\raisebox{0.3ex}{$\scriptstyle\copyright$}2018
Chinese Physical Society and the Institute of High Energy Physics
of the Chinese Academy of Sciences and the Institute
of Modern Physics of the Chinese Academy of Sciences and IOP Publishing Ltd}%

\begin{multicols}{2}

\section{Introduction}\label{section1}

In recent years, a resonance-like structure named $d^{\ast}(2380)$
was observed by the WASA-at-COSY collaborations in $pn\rightarrow
d\pi^0\pi^0$, $pn\rightarrow d\pi^+\pi^-$, when they studied the ABC
effect \cite{M. Bashkanov:2009,P. Adlarson:2011}. Later, this
particle was confirmed in series of reactions, such as $pn\rightarrow
pn\pi^0\pi^0$, $pn\rightarrow pp\pi^-\pi^0$, $pd\rightarrow
{}^3\textrm{He}\pi^0\pi^0$, $pd\rightarrow
{}^3\textrm{He}\pi^+\pi^-$, etc \cite{S.
Keleta:2009,Adlarson:2012au,Adlarson:2013usl,Adlarson:2014tcn,Adlarson:2014xmp,Adlarson:2014pxj}.
The analysis of experimental data shows that $d^{\ast}$ has a
mass of 2380 MeV, a decay width of about 70 MeV, and its spin,
isospin, and parity are $I(J^P)=0(3^+)$ \cite{Clement:2016vnl}. Since
its mass is about 70 MeV higher than the $\Delta N\pi$ threshold and
about 80 MeV lower than the $\Delta\Delta$ threshold, the threshold
effect is small for this particle. Because of its non-conventional
features of a narrow width plus a large binding energy with respect
to the $\Delta\Delta$ threshold, the structure of $d^\ast$ has attracted
attention.\\

In fact, theoretical investigations for such a state started more
than 50 years ago. After the publication of the newly observed data,
many theoretical calculations with various structural models have
been carried out. Among them, two major structural models that can
basically explain all the measured data have been investigated intensively. One of the models considers an exotic compact hexaquark-dominated structure~\cite{YZYS99,M. Bashkanov:2013,F. Huang:2015,P. Shen:2015,P. Shen:2016,P. Shen:2017,z. zhang:2017},
and the other uses a quasi-molecular resonance of $\Delta N
\pi$ (or $D_{12}\pi$)~\cite{A. Gal:2013,A. Gal:2014}. However,
apart from searching for new physical quantities to distinguish
different structural assumptions, proposing a new accessible
physical process  other than nuclear reactions or scattering
processes to confirm the
existence of the $d^{\ast}$ state is extremely important.\\

It is well-known that in the strong decay of heavy quarkonium, the
heavy quark annihilates with its anti-particle and new quark-antiquark pairs are
created. For $\Upsilon$, it can decay into a wide
variety of combinations of hadrons~\cite{C. Patrignani}.
Among these hadronic decay modes, an interesting mode
is the semi-inclusive anti-deuteron ($\bar{d}$)
production processes $\Upsilon(nS)\rightarrow\bar{d}+X$.
This is because $\bar{d}(\bar{d}^\ast)$ and $d(d^\ast)$
are bottomless baryons. The observation of $\bar{d}(\bar{d}^\ast)$ in
the $\Upsilon(nS)$ decays implies the existence of $d(d^\ast)$ in the same
process. Since the mass of $\Upsilon(nS)$ is larger than the mass of
the $d^\ast\bar{d^\ast}$ pair, the phase space is large enough for
$\Upsilon(nS)$ to decay into $\bar{d}^\ast$ (or $d^\ast$) plus other
hadrons. Therefore, it is natural to think that $\bar{d}^\ast(2380)$
is quite possible to be produced in the semi-inclusive decays of
$\Upsilon(nS)$. In this paper, we will start from the
$\Upsilon(nS)\rightarrow\bar{d}+X$ decays. By fitting the decay widths of
these processes, we can fix the unknown parameters required in the
$\Upsilon \to {\bar{d}^\ast+ X}$ calculation and then estimate the
decay widths (and/or
numbers of events) of the latter decay process.\\

The paper is organized as follows. In Section~2, the
formulism for the decay widths of the $\Upsilon(nS)\rightarrow
\bar{d}(\bar{d}^\ast)+X$ processes is briefly introduced. Numerical
results and discussions are presented in Section~3.
Finally, a short summary is given in Section~4.\\

\section{Brief formulism}\label{section2}

In this section, we will study the decay widths for
$\Upsilon(nS)\rightarrow\bar{d}^\ast(2380)+X$. With the crossing
symmetry, we investigate the scattering between $d^\ast(2380)$ and
$\Upsilon(nS)$. By virtue of the optical theorem, the
decay widths can be obtained by calculating the imaginary part of the
forward scattering amplitudes of $d^\ast(2380) + \Upsilon(nS) \rightarrow
d^\ast(2380) + \Upsilon(nS)$. In these specific elastic
scattering processes, because the quarks in $d^\ast$ and $\Upsilon$
belong to a completely different type, the exchange of quarks from
different hadrons cannot occur, while the exchange of gluon between
the quarks from different hadrons also does not contribute. The only
possible interaction between $d^\ast$ and $\Upsilon$ is the $s$-channel
meson exchange. Therefore, in the study of the
$d^\ast(d)$-$\Upsilon(nS)$ scattering, we adopt a constituent quark
model with a meson exchange potential, where we assume that the
B-meson exchange dominates, to calculate the elementary process
$\bar{b}+u(d)\rightarrow \bar{b}+u(d)$. The relevant
Feynman diagram is shown in Fig.~\ref{Fig:1}.

\begin{center}
\includegraphics [width=6cm]{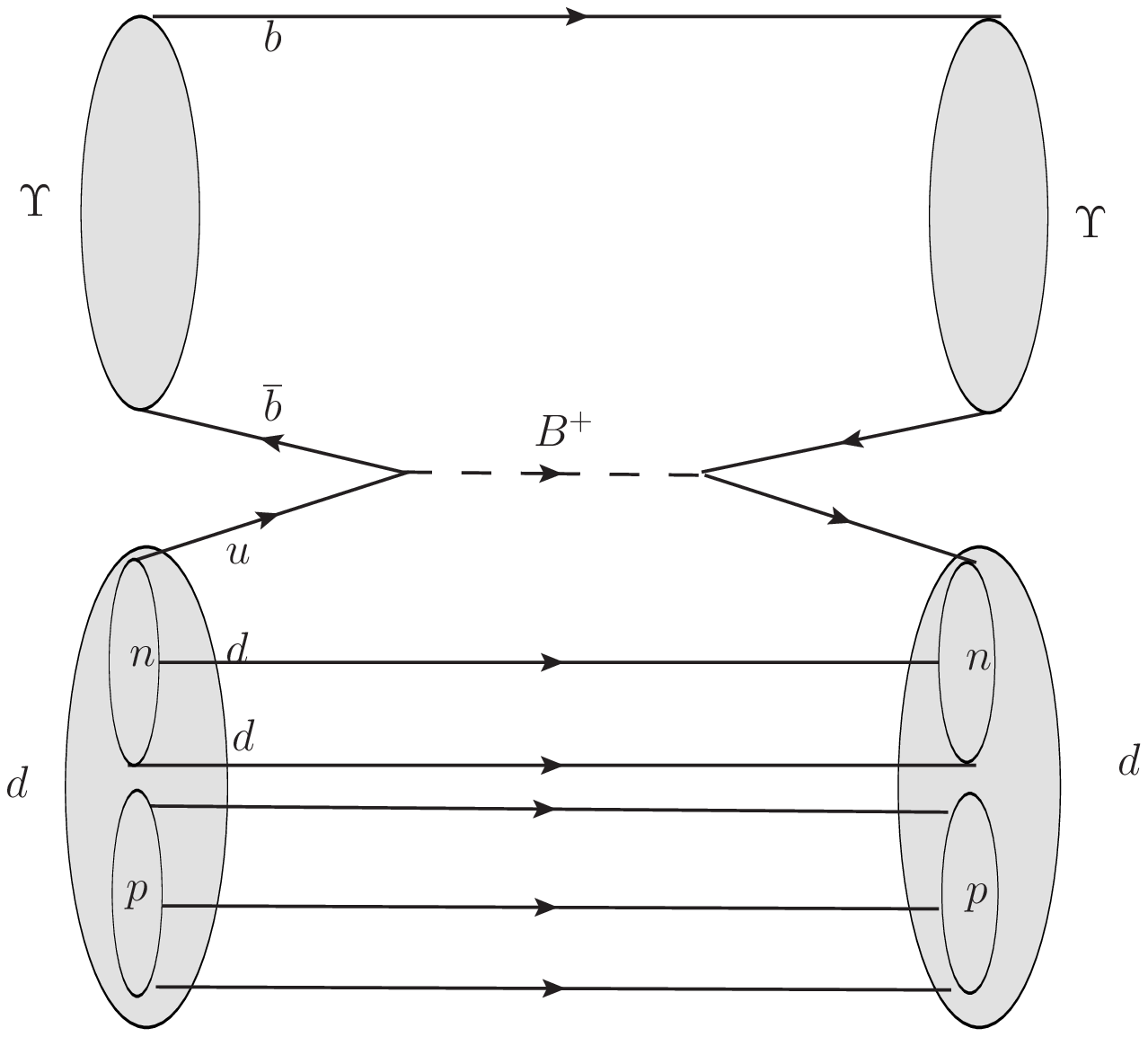}
\figcaption{\label{Fig:1}The Feynman diagram of $\Upsilon+ d\rightarrow \Upsilon+d$
forward scattering.} \label{Fig:1}
\end{center}

In this figure, the interaction between $\Upsilon$ and $d(d^\ast)$ is
governed by the quark-meson interaction. The corresponding
Lagrangian can be written as
\begin{equation} \label{eq:lag}
{\cal L}=~ig_{\bar{b}q B}\bar{\psi}_{\bar{b}}\gamma_5\psi_qB,
\end{equation}
where $g_{\bar{b}qB}$ is the coupling constant, and
$\psi_q$, $\psi_{\bar{b}}$, and $B$ denote the fields of the light quark
$q$, anti-bottom quark $\bar{b}$ and $B$ meson, respectively. To
restrict the value of the coupling constant $g_{\bar{b}q B}$, we
first study the forward scattering of $d + \Upsilon \rightarrow d +
\Upsilon$ (Fig.~\ref{Fig:1}). The scattering amplitude can be
written as
\begin{eqnarray} \label{eq:ampl}
\begin{split}
{\cal M}=&g_{\bar{b}q
B}^2\Psi^\ast_\Upsilon\Psi^\ast_d\bar{u}(p_u)\gamma_5
v(p_{\bar{b}})\dfrac{1}{q^2-m_B^2+im\Gamma}\\\nonumber
&\times\bar{v}(p_{\bar{b}})
\gamma_5u(p_u)\Psi_d\Psi_\Upsilon,
\end{split}
\end{eqnarray}
where $\Psi_\Upsilon$ and $\Psi_d$ are the wave functions of the
$\Upsilon(nS)$ and deuteron, respectively. $u(p_u)$, $\bar{u}(p_u)$,
$\bar{v}(p_{\bar{b}})$ and $v(p_{\bar{b}})$ represent the spinors of
the $u(d)$-quark and $\bar{b}$-quark in the initial and final
states, respectively. The wave functions of $\Upsilon(nS)$ can be
obtained by solving the Schr\"{o}dinger equation with a Cornell
potential \cite{E. Eichten:1980,Y. Ding:1999,Y. Lu:2016,Y. Lu:2017}.
The obtained masses for the $1S$, $2S$ and $3S$ states are 9.46 GeV,
10.02 GeV and 10.34 GeV, respectively, which are quite close to the
experimental data.\\

In the quark cluster model,
the wave function of the deuteron in the quark degrees of freedom can simply be expressed as
\begin{eqnarray} \label{eq:wfd}
\begin{split}
\Psi_{d}=&{\cal A}~[~\phi_{N}(\vec{\rho}_1,\vec{\lambda}_1)~\phi_{N}(\vec{\rho}_2,\vec{\lambda}_2)~
\eta_{NN}^{l=0}(\vec{R})~\zeta_{d}~]_{(SI)=(10)}\\
\end{split}
\end{eqnarray}
where ${\cal A}$ is the total anti-symmetrization
operator, $\phi_N$ is the internal wave function of nucleon, $\eta^{l=0}_{NN}$ represents the relative wave function in $S$-wave, which is determined by the dynamical calculation of the system with the (extended) chiral SU(3) constituent quark model,
and $\zeta_{d}$ stands for
the spin-isospin wave function in the hadronic degrees of freedom
(more details can be found in Ref.~\cite{FP:2017}). 
As commonly used, the internal wave function of the nucleon can be taken
as
\begin{eqnarray}
\phi_N=~\dfrac{1}{\sqrt{2}}[~\chi_\rho~\psi_\rho
       +~\chi_\lambda~\psi_\lambda]~\Phi_N(~\vec{\rho},~\vec{\lambda}),
\label{eq:wfn}
\end{eqnarray}
with $\chi_\rho$ ($\chi_\lambda$) and $\psi_\rho$ ($\psi_\lambda$)
being the symmetric (anti-symmetric) wave functions in the spin and
isospin spaces, $\Phi(\vec{\rho},\vec{\lambda})$ the spatial wave
function and $\rho$ and $\lambda$ the Jacobi coordinates.
In the same way, the wave function of $d^\ast$ can be
abbreviated to the form
\begin{eqnarray}\label{eq:wfd*}
\begin{split}
\Psi_{d^\ast}&={\cal A}~[~\phi_\Delta(\vec{\rho}_1,\vec{\lambda}_1)~
\phi_\Delta(\vec{\rho}_2,\vec{\lambda}_2)~\eta^{l=0}_{\Delta\Delta}(\vec{R})~
\zeta_{\Delta\Delta}~ \\
&+~\phi_{C_8}(\vec{\rho}_1,\vec{\lambda}_1)~
\phi_{C_8}(\vec{\rho}_2,\vec{\lambda}_2)~\eta^{l=0}_{C_8C_8}(\vec{R})~\zeta_{C_8C_8}]_{(SI)=(30)},
\end{split}
\end{eqnarray}
where ${\cal A}$ is the total anti-symmetrization
operator, $\phi_{\Delta}$ and $\phi_{C_8}$ denote the inner cluster wave
functions of $\Delta$ and $C_8$ (color-octet particle) in the
coordinate space, $\eta^{l=0}_{\Delta\Delta}$ and $\eta^{l=0}_{C_8C_8}$ represent
the $S$-wave relative wave functions between $\Delta\Delta$ and $C_8C_8$ clusters (the $D$-wave components are negligibly small), and $\zeta_{\Delta\Delta}$, $\zeta_{C_8C_8}$ stand for the
spin-isospin wave functions in the hadronic degrees of freedom in
the corresponding channels, respectively.
The channel wave function can be defined as
\begin{equation}
\chi^{eff,l=0}_{\Delta\Delta(C_8C_8)}(\vec{R})=<\phi_{\Delta(C_8)}(\vec{\rho}_1,\vec{\lambda}_1)~
\phi_{\Delta(C_8)}(\vec{\rho}_2,\vec{\lambda}_2)|\Psi_{d^\ast}>.
\end{equation}
Therefore, for simplicity, the wave function of $d^*$ can be rewritten as
\begin{eqnarray}
\begin{split}
\Psi_{d^\ast} \cong &[~\phi_\Delta(\vec{\rho}_1,\vec{\lambda}_1)~
\phi_\Delta(\vec{\rho}_2,\vec{\lambda}_2)~\chi^{eff,l=0}_{\Delta\Delta}(\vec{R})~
\zeta_{\Delta\Delta}~ \\ \nonumber
&+~\phi_{C_8}(\vec{\rho}_1,\vec{\lambda}_1)~
\phi_{C_8}(\vec{\rho}_2,\vec{\lambda}_2)~\chi^{eff,l=0}_{C_8C_8}(\vec{R})~\zeta_{C_8C_8}]_{(SI)=(30)}.
\end{split}
\end{eqnarray}
We should emphasize that our treatment for the wave function of $d^*$ is just an approximation for
reducing tedious and almost inoperable calculations and making the two components orthogonal.
The obtained effective relative wave function can reasonably contain most of the effect of anti-symmetrization
of the wave function of $d^*$ shown in Eq.~(\ref{eq:wfd*}).
The effective wave function is obtained by projecting onto the physical base, and further described by the sum of four Gaussian functions.
\begin{eqnarray}
\label{eq:chwf}
\begin{split}
\chi^{eff,l=0}_{\Delta\Delta(C_8C_8)}(\vec{R})=~\sum_{i=1}^{4}~c_i~
\textrm{exp}(-\dfrac{\vec{R}^2}{2b_i^2}).
\end{split}
\end{eqnarray}
It should be noticed that
the quantum numbers for the color-octet $C_8$-cluster and the
color-singlet $\Delta$-cluster in $d^\ast$ are different. For the
$C_8$-cluster, $S=3/2, I=1/2, C=(11)$, while for the $\Delta$-cluster,
$S=3/2, I=3/2, C=(00)$, where $S$, $I$ and $C$ denote the spin,
isospin and color, respectively. It should be specially mentioned
that these two channel wave functions are orthogonal to each other
and contain all the effects of the totally anti-symmetrization
implicitly. The details can be found in Refs.~\cite{F. Huang:2015,P.
Shen:2015,P. Shen:2016,z. zhang:2017,FP:2017}.
\\

On the other hand, the data published by the PDG~\cite{C. Patrignani}
show that the width of the $B$ meson is very small. Therefore, the
propagator can be written as
\begin{equation}\label{eq:propa}
\dfrac{1}{q^2-m_B^2+im\Gamma}\simeq~\frac{1}{q^2-m_B^2+i\epsilon}\rightarrow~
-2\pi i\delta(q^2-m_B^2).
\end{equation}
Then, the imaginary part of the amplitude is expressed as


\begin{eqnarray}\label{eq:im}
\begin{split}
\mathrm{Im}\mathcal{M}=-\int d\,\Pi\ 2\pi g_{\bar{b}q B}^2
\Psi^\ast_\Upsilon\Psi^\ast_d\bar{u}(p_u)\gamma_5v
(p_{\bar{b}}) \\
\bar{v}(p_{\bar{b}})\gamma_5u(p_u)
\Psi_d\Psi_\Upsilon\delta(q^2-m_B^2),
\end{split}
\end{eqnarray}
where $d\,\Pi$ is an integral measure, including $d\vec{p}_\rho$,
$d\vec{p}_\lambda$, $d\vec{p}_\eta$ and $d\vec{p}_R$ for the internal momenta in the
nucleon, internal momentum of $\Upsilon$ and relative momentum between
nucleons, respectively. The momenta $\vec{p}_{\bar{b}}$, $\vec{p}_u$ and
$\vec{q}=\vec{p}_{\bar{b}}+\vec{p}_u$ are related to these momenta and the momentum
$\vec{p}_d$ of the deuteron through the Jacobian transformation shown in Appendix A.
The forward scattering condition requires that the momenta and quantum numbers
of quarks in the initial and finial states do not change.\\

With the optical theorem, the semi-inclusive decay width of
$\Upsilon\rightarrow \bar{d} +X$ can be calculated by
\begin{equation}
\label{eq:width}
\textrm{d}\Gamma=
\dfrac{1}{m_\Upsilon}(\frac{\textrm{d}\vec{p}_{d}}{(2\pi)^{3}}\dfrac{1}{2E_{d}})
\times \mathrm{Im}\mathcal{M}(\Upsilon+d\rightarrow \Upsilon+d).
\end{equation}
Integrating over the possible range of $p_d$, the decay
widths of $\Upsilon(nS)$ to $\bar{d}+X$ can be obtained. In the final
step, the upper limit of the $p_d$ integration is determined by the
four-momentum conservation
\begin{equation}
\label{eq:econs}
\sqrt{\vec{p}_d^{~2}+m_d^{2}}+\sqrt{\vec{p}_d^{2}+M_X^2}=m_\Upsilon.
\end{equation}
with $M_X$ being the residual mass of all final
particles in this semi-inclusive decay except the deuteron. For the
semi-inclusive decays of $\Upsilon(nS)$ to $\bar{d}^\ast(2380)+X$, the
calculation can be carried out in the same way, but replacing the
wave function and the mass of the deuteron with those of the
$d^\ast(2380)$.

\end{multicols}

\begin{figure*}[!htbp]
\centering
\includegraphics [width=5.5cm, height=4.5cm]{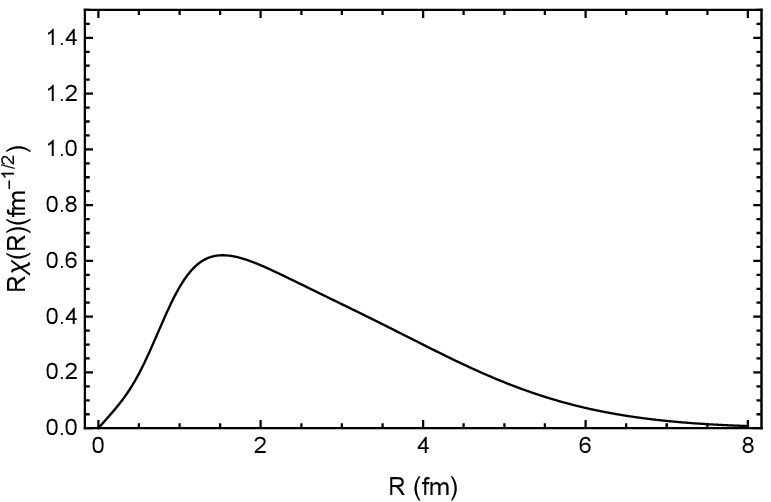}
{\hskip 0.25cm}
\includegraphics [width=5.5cm, height=4.5cm]{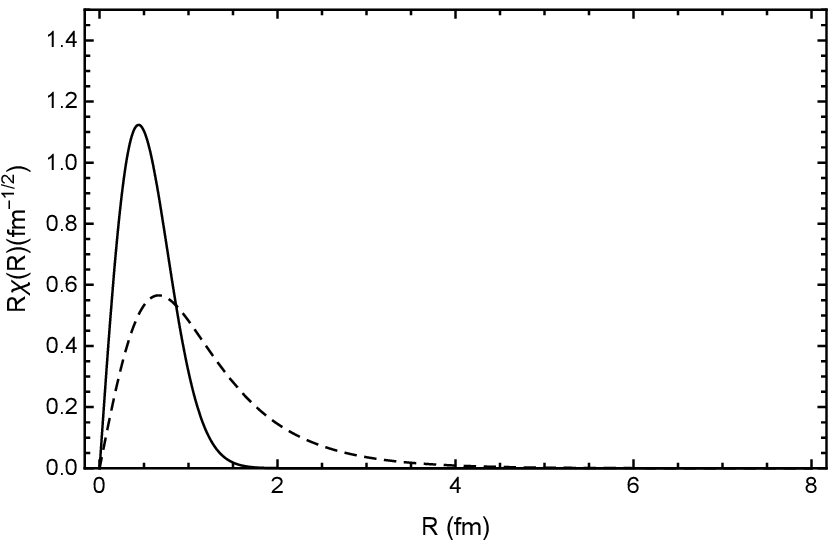}
\caption{Channel wave functions in the $S$ partial wave by using the
extended chiral SU(3) quark model. The solid curve in the left-hand
diagram shows the wave function of the deuteron, and the dashed and
solid curves in the right-hand diagram show the wave functions of the $\Delta\Delta$ and
$C_8C_8$ components of $d^\ast(2380)$, respectively.} \label{fig:wf}
\end{figure*}
\begin{multicols}{2}

\section{Numerical results and discussion}\label{section3}

In this section, we will present the numerical results for the widths
of the semi-inclusive decays $\Upsilon(nS)\rightarrow
\bar{d}(\bar{d}^\ast)+X$. Before calculating the widths,
we present the channel wave functions of deuteron and
$d^\ast(2380)$ (both $\Delta$$\Delta$ and $C_8C_8$
components) in Fig.~\ref{fig:wf}.

In this figure, we only plot the wave functions in the $S$ partial
wave and ignore those in the $D$ partial wave, because the latter is
negligibly small. From these curves, one clearly sees that the size
of $d$ is larger than that of $d^\ast$, and the size of the $C_8C_8$
component is even smaller. The peaks of the wave functions for the
deuteron, the $\Delta\Delta$ and $C_8C_8$ components of $d^\ast$ are
located around $1.5$ fm, $0.8$ fm and $0.5$ fm, respectively.\\

With these wave functions and
Eqs.~(\ref{eq:propa})-(\ref{eq:width}), we are able to calculate the
decay widths of $\Upsilon(nS)$ to $\bar{d}(\bar{d}^\ast)+X$.
We firstly calculate the widths of $\Upsilon(nS)$
decaying to $\bar{d}+X$ and compare them with the data to restrict
the value of the coupling constant $g_{\bar{b}q B}$. As mentioned
above, due to energy conservation, the maximal
momentum of the deuteron should satisfy
Eq.~(\ref{eq:econs}), namely it is residual mass dependent. However,
in the semi-inclusive decay, the only affirmed information for $X$
is its baryon number being 2, and the mass of $X$ may have a value
within a broad range of about $2-7$ GeV. Therefore, the maximal value of
$p_d$, and consequently the resultant decay widths, will vary
according to the residual mass of $M_X$, or somehow relate to the
momentum transfer $\bm q$ (more momentum related content will be
discussed later). To get a meaningful result, we treat this issue by
inserting a phenomenological form factor of the $p_d$ distribution:
\begin{equation}
F(p_d) ={\cal N} \text{exp}\big [-\frac{(p_d-p_0)^2}{\Lambda^2}\big ],
\end{equation}
where ${\cal N}$ denotes the normalization factor, $p_0$ is the
most probable distribution point of $p_d$, and $\Lambda$
describes the degree of the $p_d$ extension. Therefore, in our
calculation, there are three parameters, $p_0$, $\Lambda$, and the
coupling constant $g_{\bar{b}qB}$. $p_0$ and $\Lambda$ are determined by the
experimental momentum distribution of the $\bar{d}$ from the semi-inclusive processes
$\Upsilon(nS) \rightarrow \bar{d}+X$. $g_{\bar{b}qB}$ is determined by the experimental decay width data.

\end{multicols}

\begin{figure}[!htbp]
\centering
\includegraphics [width=5cm, height=4cm]{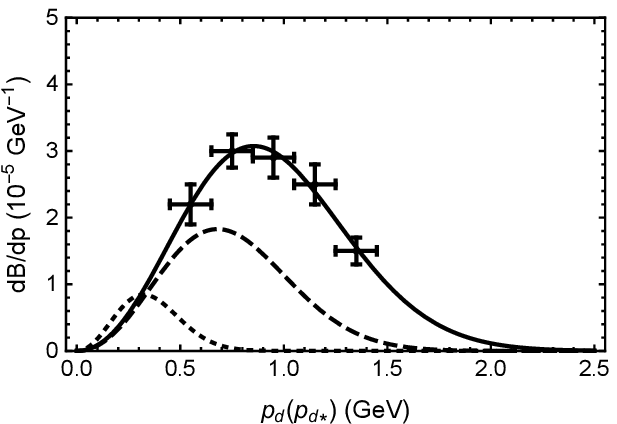}
{\hskip 0.4cm}
\includegraphics [width=5cm, height=4cm]{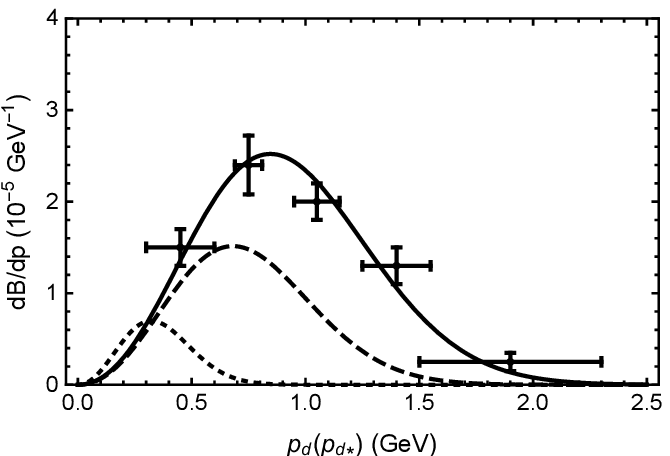}
{\hskip 0.4cm}
\includegraphics [width=5cm, height=4cm]{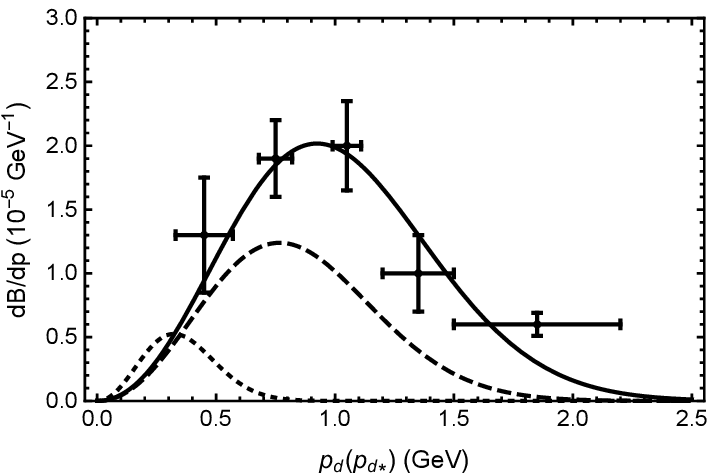}
{\hskip 0.4cm}
\caption{The momentum distributions of $\bar{d}$ ($\bar{d}^\ast$) for the process
$\Upsilon(nS) \rightarrow \bar{d} (\bar{d}^\ast) + X$. The figures, from left to right, are for
decays of $\Upsilon(1S)$, $\Upsilon(2S)$ and $\Upsilon(3S)$, respectively. The dots are data observed in experiments. The solid lines are fits for $\bar{d}$ and the dashed lines are predictions for $\bar{d}^\ast$.}

\label{Fig:3}
\end{figure}

\begin{multicols}{2}

Inserting the above form factor into the original Eq.~(\ref{eq:width}), we can get the final momentum
distribution of $\bar{d}$. In Fig.~\ref{Fig:3}, the momentum distribution of $\bar{d}$ is shown by
solid lines. The figures, from left to right, are for the decay of  $\Upsilon (1S)$, $\Upsilon (2S)$
and $\Upsilon (3S)$, respectively. For example, for $\Upsilon (1S)$,
$p_0$ and $\Lambda$ are determined to be 0.24 GeV and 0.8 GeV, which provide the best fit to
the experimental momentum distribution \cite{CLEO}. With the obtained $p_0$ and $\Lambda$,
$g_{\bar{b}qB}$ is determined to be $2.5 \times 10^{-3}$
to get the decay width of $\Upsilon(1S) \rightarrow \bar{d} + X$ $ 154 \times 10^{-5}$ KeV.
For $\Upsilon (2S)$ and $\Upsilon (3S)$, from the experimental distributions \cite{babar},
$p_0$/$\Lambda$ are obtained as 0.25 GeV/0.78 GeV and 0.27 GeV/0.87 GeV, respectively.

It should be mentioned that in our model calculation,
$g_{\bar{b}qB}$ is considered phenomenologically as an effective
coupling constant, instead of a momentum dependent one. In other
words, we neither use a running coupling constant nor
add a form factor to the vertex of the quark-meson
interaction. Since the mass of $\Upsilon(nS)$ increases
with increasing main quantum number $n$, the momentum dependence of
their semi-inclusive decay widths will also vary, namely the
phenomenological coupling constants for different $\Upsilon(nS)$
states should have certain deviations. To compensate for this
difference, we determine the $g_{\bar{b}qB}$ for different
$\Upsilon(nS)$ states by fitting their own observed decay widths.
As a result, the obtained effective coupling constants
$g_{\bar{b}qB}(nS)$ for $\Upsilon(2S)$ and $\Upsilon(3S)$ are
$1.9\times10^{-3}$ and $1.3\times10^{-3}$, respectively. This is
consistent with the result from the form factor method, where the
effective coupling constant decreases with the increasing
momentum.\\

With the determined effective coupling constants for corresponding
$\Upsilon(nS)$, we can proceed with the calculations for the decay widths
of the $\Upsilon(nS) \rightarrow \bar{d}^\ast(2380) +X$ processes. In
the calculation, $g_{\bar{b}qB}(nS)$ for different
$nS$ states take the same values as those in the corresponding
deuteron case. However, we have no information on the momentum distributions
of the $\bar{d}^\ast$, i.e., we do not know the exact values of
$p_0$ and $\Lambda$. Since the mass of $d^\ast$ is larger than that of the deuteron,
one can imagine that $p_0$ for the $\bar{d^\ast}$ is smaller than that for $\bar{d}$.
Therefore, in our calculation, $p_0$ and $\Lambda$ in the $\bar{d^\ast}$ case
are chosen properly to be in relatively large ranges. For example,
for $\Upsilon(nS)$, $p_0$ is chosen to be in the range $0.05 - 0.2$ GeV,
which is smaller than that in the anti-deuteron case. The range of $\Lambda$ is chosen to be $0.3 - 0.6$ GeV
for $\Upsilon(1S)$ and $\Upsilon(2S)$ and $0.3 - 0.7$ GeV for $\Upsilon(3S)$. Similar to the $\bar{d}$ case,
for $\bar{d^\ast}$, the final momentum distributions are obtained by inserting the
form factors of $\bar{d^\ast}$ into Eq.~(\ref{eq:width}).
The resultant momentum distributions are shown in Fig.~\ref{Fig:3} by dashed lines.
The left-hand dashed line in each figure is for $(p_0, \Lambda) = (0.05$ GeV$, 0.3$ GeV$)$ and the right-hand dashed line is for $(p_0, \Lambda)$ = (0.2 GeV, 0.6 GeV (0.7 GeV for $\Upsilon(3S)$)).
It is found that the smaller the momentum of $\bar{d}^\ast$, the lower its production rate.

\end{multicols}

\begin{figure}[!htbp]
\centering
\includegraphics [width=5cm, height=4cm]{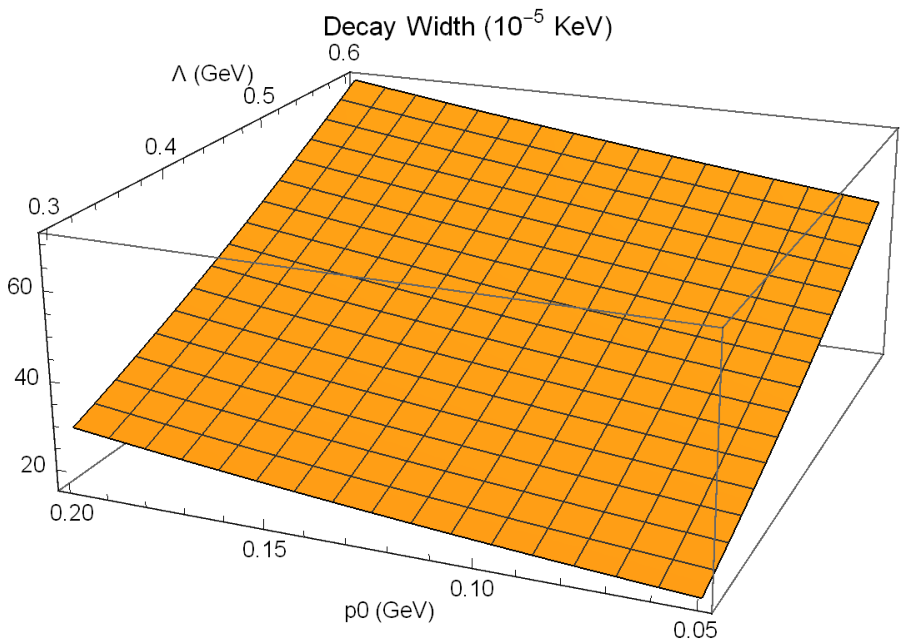}
{\hskip 0.4cm}
\includegraphics [width=5cm, height=4cm]{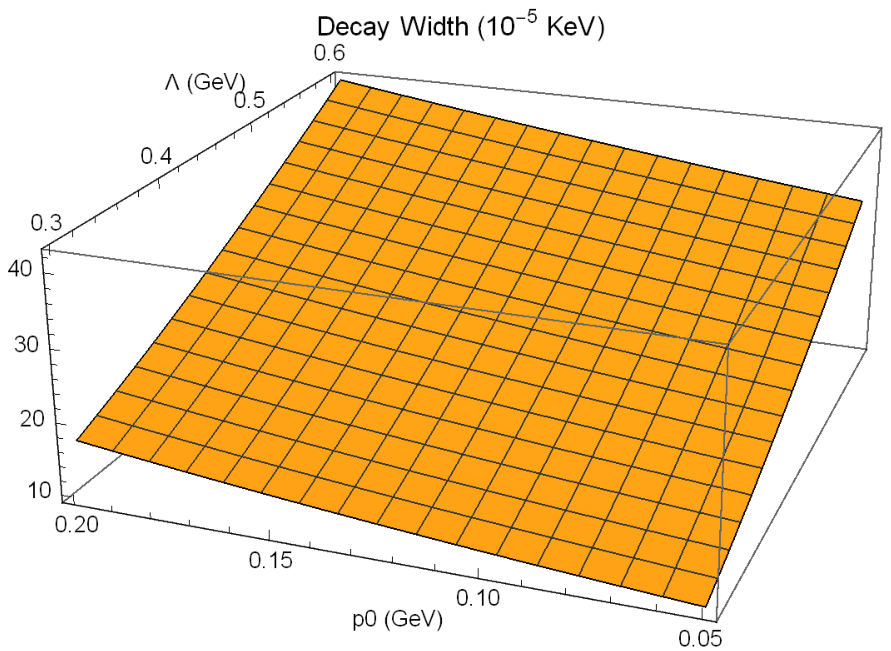}
{\hskip 0.4cm}
\includegraphics [width=5cm, height=4cm]{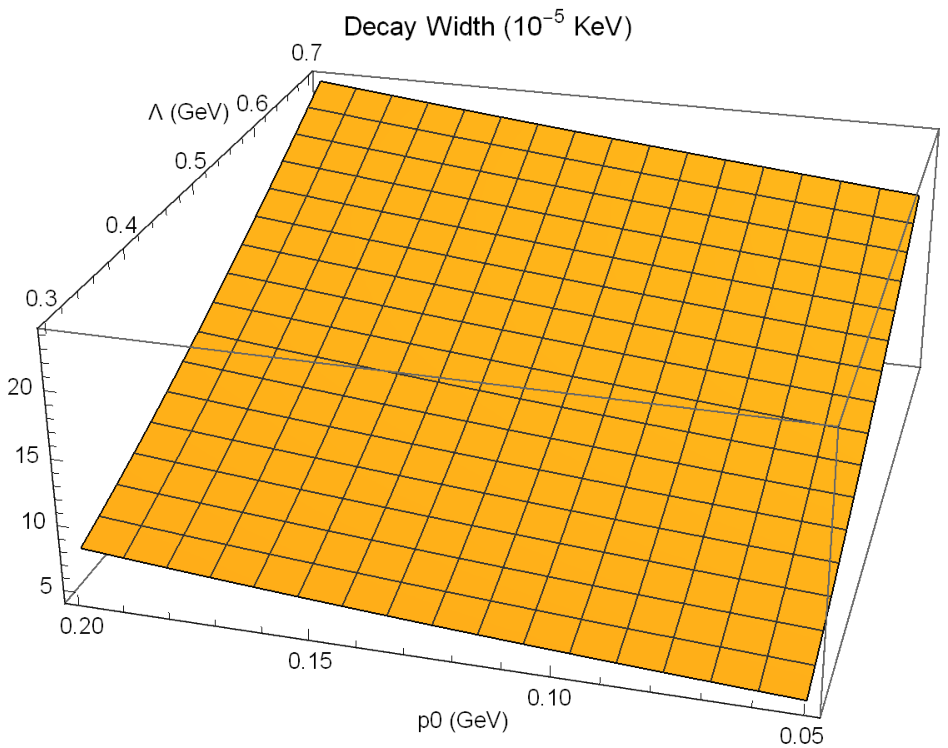}
{\hskip 0.4cm}
\caption{The predicted decay widths of $\Upsilon(nS) \rightarrow
\bar{d}^\ast + X$ versus $p_0$ and $\Lambda$. The figures, from left to right,
are for $\Upsilon(1S)$, $\Upsilon(2S)$ and $\Upsilon(3S)$, respectively.}
\label{Fig:4}
\end{figure}

\begin{multicols}{2}

In Fig.~4, we show the decay widths of $\Upsilon(nS) \rightarrow
\bar{d}^\ast + X$ versus $p_0$ and $\Lambda$ using 3-dimensional figures.
The figures, from left to right, correspond to $\Upsilon(1S)$, $\Upsilon(2S)$
and $\Upsilon(3S)$, respectively. From the figure, one can see that the decay widths increase with increasing $p_0$ and $\Lambda$. For $\Upsilon(1S)$, the predicted range of width
is from $16 \times 10^{-5}$ keV to $71 \times 10^{-5}$ keV.
The smallest number is for $p_0 = 0.05$ GeV and $\Lambda=0.3$ GeV, while the
largest number is for $p_0 = 0.2$ GeV and $\Lambda=0.6$ GeV. Compared with
the decay width of $\Upsilon(1S) \rightarrow
\bar{d} + X$, the production rate of $\bar{d}^\ast$ is suppressed by about
$2 - 10$ times. For $\Upsilon(2S)$ and $\Upsilon(3S)$,
the widths are in the ranges from $9.6 \times 10^{-5}$ keV to $42 \times 10^{-5}$ keV
and $4.4 \times 10^{-5}$ keV to $24 \times 10^{-5}$ keV,  which are
also about $2-10$ times smaller than the corresponding cases of $\bar{d}$.
The production rates of $\bar{d}^\ast$ are smaller than $\bar{d}$,
mainly because of the mass difference between $\bar{d}^\ast$ and
$\bar{d}$. The difference of the wave functions has only a slight
influence, though the shapes of their wave functions are quite
different.\\

The calculated momentum of $\bar{d^\ast}$ is likely in the range $0.3 - 0.8$ GeV.
Therefore, if we want to find $\bar{d^\ast}$ in $\Upsilon (nS)$ decays, we need to detect it in this momentum region.
With the chosen parameter ranges, the production rate of $\bar{d^\ast}$ is suppressed by $2-10$ times compared with
the corresponding production rate of $\bar{d}$. Finding $\bar{d^\ast}$ is still within experimental ability~\cite{exp}.
However, if the momentum of $\bar{d^\ast}$ is concentrated in an even smaller region due to some special mechanism,
$\bar{d^\ast}$ will be hard to discover from $\Upsilon (nS)$ decays at  current experimental facilities.

\end{multicols}

\begin{center}
\tabcaption{\label{tab:widthds}The coupling constants, parameters of the momentum distribution, and decay widths for the process of $\Upsilon(nS)\rightarrow $$\bar{d}$ ($\bar{d}^\ast$(2380))+X.}
\footnotesize
\begin{tabular*}{170mm}{@{\extracolsep{\fill}}cccccccc}
\toprule
{State} & $g_{\bar{b}qB} (10^{-3})$ & $p_0^{\bar{d}}$ (GeV) & $\Lambda^{\bar{d}}$ (GeV) & $\Gamma_{\bar{d}}$ ($10^{-5}$ KeV)
& $p_0^{\bar{d}^*}$ (GeV) & $\Lambda^{\bar{d}^*}$ (GeV) & $\Gamma_{\bar{d}^\ast}$ ($10^{-5}$ KeV) \\
\hline
$1S$ & 2.5 & 0.24 & 0.80 & $154$ & $0.05-0.2$ & $0.3-0.6$ & $16-71$   \\

$2S$ & 1.9 & 0.25 & 0.78 & $89$ & $0.05-0.2$ & $0.3-0.6$ & $9.6-42$  \\

$3S$ & 1.3 & 0.27 & 0.87 & $47$ & $0.05-0.2$ & $0.3-0.7$ & $4.4-24$ \\
\bottomrule
\end{tabular*}
\end{center}

\begin{multicols}{2}

\section{Summary}\label{section4}

We calculated the widths of semi-inclusive decays of $\Upsilon(nS)
\rightarrow \bar{d}^\ast(2380)+ X$. With the help of crossing
symmetry, the decay widths are obtained by investigating the imaginary
part of the forward scattering amplitudes between $d^\ast$ and
$\Upsilon(nS)$. The wave functions of the deuteron and $d^\ast$ are
obtained from  chiral SU(3) quark model calculations,
and the wave functions of $\Upsilon(nS)$ are calculated
by solving the Schr\"{o}dinger equation with a Cornell potential. In
the $\Upsilon(nS)-d^\ast$ scattering, as a rough estimation, a $s$-channel $B$-meson exchange is
assumed as a dominant interaction where the basic coupling constant
$g_{\bar{b}qB}$ is phenomenologically determined by fitting the
observed partial decay widths of the $\Upsilon(nS)\rightarrow
\bar{d}+X$ processes. To compensate for the lack of information about X, a Gaussian function is introduced
and $p_0$ and $\Lambda$ in the function are obtained by reproducing
the experimental momentum distributions of $\bar{d}$.
For the $\bar{d^\ast}$ cases, since we have no information on the momentum distribution
of $d^\ast$ in the final states, we study $d^\ast$ in relatively large ranges for
$p_0$ and $\Lambda$. The calculated momentum of the generated $\bar{d^\ast}$ is in the range $0.3-0.8$ GeV.
The overall widths of
$\Upsilon(nS)~(n=1,2,3)$ decaying into $\bar{d}^\ast+X$
are about $(16-71) \times 10^{-5}$ keV, $(9.6-42)
\times 10^{-5}$ keV, and $(4.4-24) \times 10^{-5}$ keV, respectively,
which are about $2-10$ times smaller
than those in the $\bar{d}$ cases.
It is still within experimental ability to find $\bar{d^\ast}$ in  $\Upsilon (nS)$ decays.
The difference in the wave
functions between deuterons and $d^\ast$ has only  a tiny influence on the
productions of $\bar{d}$ and $\bar{d}^\ast$. The suppression of the
production rates of $\bar{d}^\ast$ is mainly due to the
mass difference (or phase space difference) between $\bar{d}$ and $\bar{d}^\ast$.
We summarize our obtained values of the coupling constants, parameters for momentum distribution, and decay widths
in Table 1. The labels $\bar{d}$ and $\bar{d^\ast}$ are shown explicitly for these two cases.
As a conclusion, it is very likely that one can find $\bar{d}^\ast(2380)$ in the semi-inclusive decays of $\Upsilon(nS)$.

\section{Acknowledgments}

\acknowledgments{The authors thank F. Huang for providing the wave functions
of the $d^\ast$ and deuteron, and thank Z. X. Zhang, C. Z. Yuan and C. P. Shen for helpful discussions.
Chao-Yi L\"{u} is grateful to Yu Lu for suggestion on Mathematica.}

\end{multicols}

\vspace{10mm}

\begin{multicols}{2}

\subsection*{Appendices A}
\begin{small}

Jacobian transformations for the $\Upsilon$, nucleon and deuteron systems, respectively.
\begin{equation}
\begin{cases}
\vec{p}_{\eta}=\dfrac{1}{2}(\vec{p}_b-\vec{p}_{\bar{b}})\\
\vec{k}=(\vec{p}_b+\vec{p}_{\bar{b}}),
\end{cases}
\end{equation}
\begin{equation}
\begin{cases}
\vec{p}_{\rho}=\dfrac{1}{2}(\vec{p}_{u}-\vec{p}_{d_1})\\
\vec{p}_{\lambda}=\dfrac{1}{3}(\vec{p}_u+\vec{p}_{d_1}-2\vec{p}_{d_2})\\
\vec{p}_n=(\vec{p}_u+\vec{p}_{d_1}+\vec{p}_{d_2}),
\end{cases}
\begin{cases}
\vec{p}_{R}=\dfrac{1}{2}(\vec{p}_n-\vec{p}_p)\\
\vec{p}_{d}=(\vec{p}_n+\vec{p}_p),
\end{cases}
\end{equation}
where $\vec{p}_{\rho}$, $\vec{p}_{\lambda}$ and $\vec{p}_{\eta}$ are the internal momenta of the nucleon, and internal momentum of $\Upsilon$, respectively. $\vec{p}_n$ and $\vec{k}$ represent the momenta of the nucleon and $\Upsilon$, respectively.
$\vec{p}_{R}$ denotes the relative momentum between nucleons, and $\vec{p}_{d}$ is the center of mass momentum of the deuteron.
The momenta on the right-hand side of the equations are those of the particles labeled by the subscripts.

\end{small}

\end{multicols}

\vspace{-1mm}
\centerline{\rule{80mm}{0.1pt}}
\vspace{2mm}

\begin{multicols}{2}

\end{multicols}

\clearpage

\end{CJK*}
\end{document}